# Bi$_2$Te$_3$/Sb$_2$Te$_3$ Heterojunction and Thermophotovoltaic Cells Absorbing the Radiation from Room-temperature Surroundings

Xiaojian Li[1], Chaogang Lou[1*], Xin Li[1], Yujie Zhang[1], Bo Yin[2]

[1]*Joint International Research Laboratory of Information Display and Visualization, School of Electronic Science and Engineering, Southeast University, Nanjing, 210096, P. R. China*
[2]*School of Physics and Electronics, Nanyang Normal University, Nanyang, 473007, P. R. China*

The thermophotovoltaic cells which can convert the infrared radiation from room-temperature surroundings into electricity are of significance due to their potential applications in many fields. In this work, narrow bandgap Bi$_2$Te$_3$/Sb$_2$Te$_3$ thin film thermophotovoltaic cells were fabricated, and the formation mechanism of Bi$_2$Te$_3$/Sb$_2$Te$_3$ p-n heterojunctions was investigated. During the formation of the heterojunctions at room temperature, both electrons and holes diffuse in the same direction from n-type Bi$_2$Te$_3$ thin films to p-type Sb$_2$Te$_3$ thin films rather than conventional bi-directional diffusion. Because the strong intrinsic excitation generates a large number of intrinsic carriers which weaken the built-in electric field of the heterojunctions, their I-V curves become similar to straight lines. It is also demonstrated that Bi$_2$Te$_3$/Sb$_2$Te$_3$ thermophotovoltaic cells can output electrical power under the infrared radiation from a room-temperature heat source. This work proves that it is possible to convert the infrared radiation from dark and room-temperature surroundings into electricity through narrow bandgap thermophotovoltaic cells.

## I. INTRODUCTION

In recent years, thermophotovoltaic cells have become attractive due to their capability to convert heat into electricity directly [1]. Different from solar cells which absorb visible and near infrared photons, the thermophotovoltaic cells can absorb long-wavelength infrared photons because they are made of narrow bandgap semiconductors. Considering the extensive application of the cells, many efforts have been paid and good progresses have been made [7-11].

So far, all of the researches in this field focus on the thermophotovoltaic cells which work well under the radiation from the heat sources with high-temperature [12-15]. It has not been reported about the thermophotovoltaic cells which can absorb the infrared photons radiated from room-temperature objects. The possible reason is that this kind of cells are made of the narrow bandgap semiconductors which have strong intrinsic excitation effect at room temperature. The effect generates a large number of intrinsic carriers which deteriorate the rectification characteristics of p-n junctions [16, 17].

Due to their potential applications, it is of significance to develop the thermophotovoltaic cells which can convert the room-temperature radiation into electricity. In this work, we have fabricated the thermophotovoltaic cells with the p-n heterojunctions consisting of n-type Bi$_2$Te$_3$ thin films and p-type Sb$_2$Te$_3$ thin films, both of which are narrow bandgap semiconductors and can absorb the infrared photons radiated from room-temperature objects. The formation mechanism of the heterojunction at room temperature is investigated, and the output electrical power of the cells is measured.

## II. EXPERIMENT

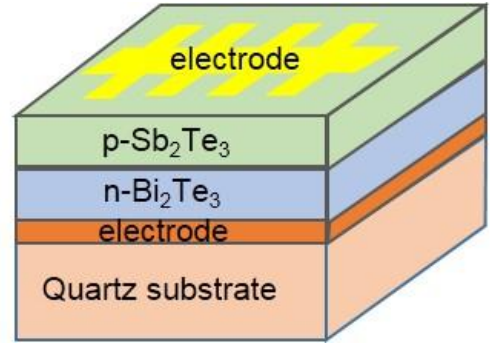

FIG. 1. The schematic structure of the narrow bandgap Bi$_2$Te$_3$/Sb$_2$Te$_3$ thermophotovoltaic cells.

The schematic structure of Bi$_2$Te$_3$/Sb$_2$Te$_3$ thermophotovoltaic cells is shown in Fig. 1. In the experiments, 100 nm Mo films were at first deposited on quartz substrates (2 cm✕2 cm) as the bottom electrode by DC magnetron sputtering. Then, Bi$_2$Te$_3$ thin films and Sb$_2$Te$_3$ thin films were deposited by RF magnetron sputtering successively. RF powers for depositing Bi$_2$Te$_3$ and Sb$_2$Te$_3$ thin films were 100 W, and the sputtering time of Bi$_2$Te$_3$ and Sb$_2$Te$_3$ thin films was set as 300 s and 400s, respectively. 100 nm Au films as the top electrode were fabricated by vacuum thermal evaporation.

---

* lcg@seu.edu.cn

The I-V curves of the heterojunction were measured at liquid nitrogen temperature and at room temperature by Keithley 2400. The carrier concentration was given by HMS-5000 Variable Temperature Hall Effect Measurement System. The compositions of $Bi_2Te_3$ thin films and $Sb_2Te_3$ thin films were measured by AMETEK PV77-60680 ME EDX.

We also tested the short-circuit current and the open-circuit voltage of $Bi_2Te_3/Sb_2Te_3$ thermophotovoltaic cells under the radiation from the heat source with different temperatures. The cells were put into a stainless steel box with an aperture of the diameter 1 cm, through which the infrared photons emitted from an outside heat source can reach the cells. A copper coil was electrically heated and used as the heat source whose temperature varies from 300 K to 470 K. During the test, the thermophotovoltaic cells is fixed at room temperature.

### III. RESULTS AND DISCUSSION

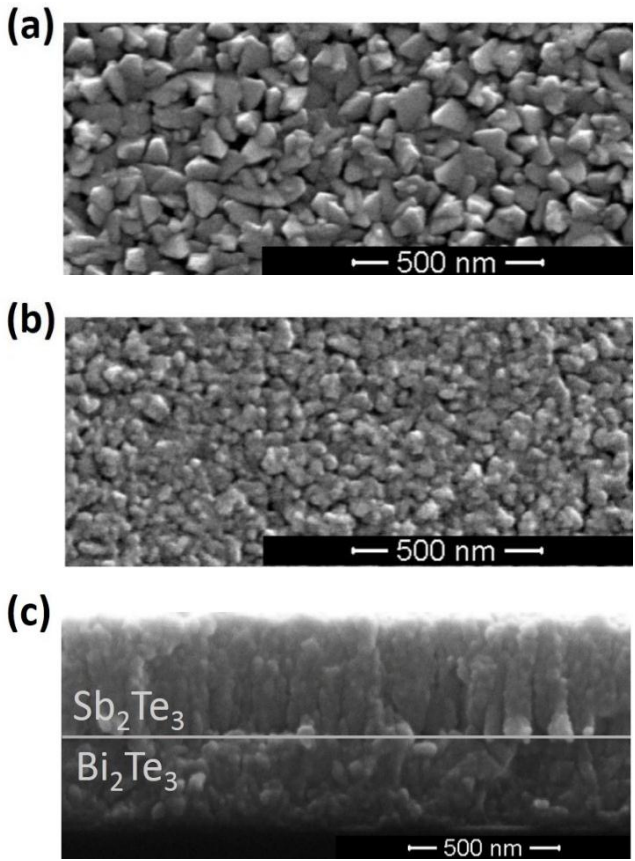

FIG. 2. SEM images of the surface morphologies of (a) $Bi_2Te_3$ thin films and (b) $Sb_2Te_3$ thin films. (c) Cross section of $Bi_2Te_3/Sb_2Te_3$ p-n heterojunction

Figure 2(a) and Fig. 2(b) are the SEM images of top view of $Bi_2Te_3$ thin films and $Sb_2Te_3$ thin films, respectively. Both are polycrystalline thin films which consist of the grains with the size from dozens of nanometers to one hundred nanometers. The atomic ratio of $Bi_2Te_3$ thin films is Bi:Te=39.20:60.80 and that of $Sb_2Te_3$ thin films is Sb:Te=40.95:59.05. Both agree with the stoichiometric composition of the thin films.

Figure 2(c) shows the cross-section of $Bi_2Te_3/Sb_2Te_3$ heterojunctions. It is a two-layer structure, and the thicknesses of $Bi_2Te_3$ thin films and $Sb_2Te_3$ thin films are about 310 nm and 380 nm, respectively.

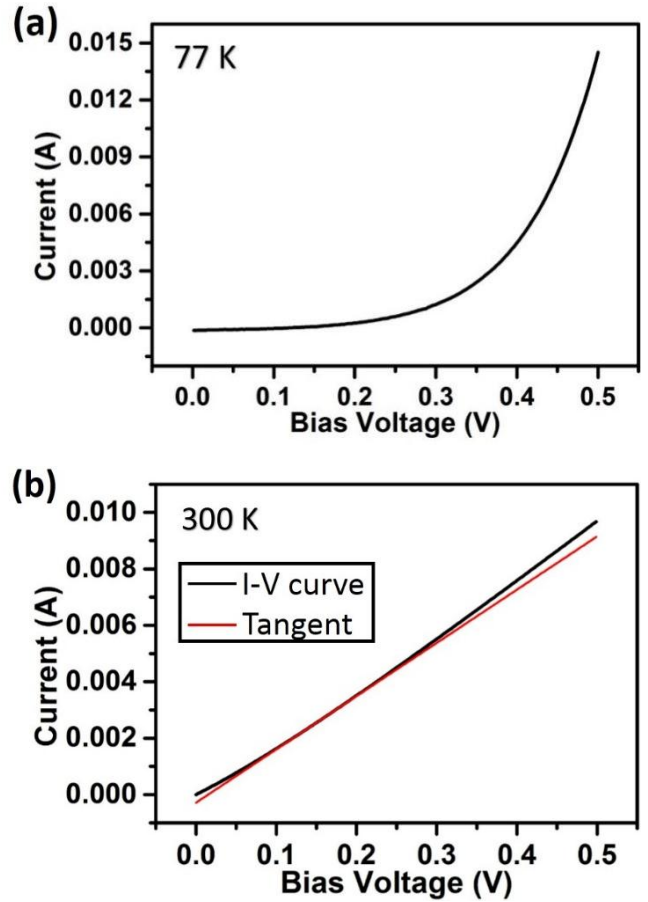

FIG. 3. I-V curves of $Bi_2Te_3/Sb_2Te_3$ heterojunction (a) at liquid nitrogen temperature (77 K) and (b) at room temperature (300 K).

Figure 3(a) shows the I-V curve of $Bi_2Te_3/Sb_2Te_3$ heterojunctions at liquid nitrogen temperature (77 K). It is a typical I-V curve which verifies the formation of the p-n junction. When the temperature rises to room temperature (300 K), the I-V curve becomes different, as shown in Fig. 3(b). For the purpose of comparison, the tangent line of the I-V curve at 0.15 V is given. It can be seen that the I-V curve becomes similar to a straight line. This indicates that, at room temperature, $Bi_2Te_3/Sb_2Te_3$ heterojunctions are more similar to a resistor than to a p-n junction.

This phenomenon can be attributed to the intrinsic excitation in $Bi_2Te_3/Sb_2Te_3$ heterojunction. At liquid nitrogen temperature, the intrinsic excitation is weak and the intrinsic carrier concentration is negligible. In this case, the p-n heterojunctions are formed through the bi-directional diffusion of the majority carriers and have a typical I-V curve.

However, at room temperature, the intrinsic excitation becomes strong and the intrinsic carrier concentration can not be neglected. The p-n junctions are formed through the carriers' diffusion in the same direction instead of the conventional bi-directional diffusion. The followings are the detailed explanation.

Table I. Majority carrier concentrations of n-type $Bi_2Te_3$ (electron) and p-type $Sb_2Te_3$ (hole) at 77 K and 300 K

|  | Majority Carrier Concentration($cm^{-3}$) | |
| --- | --- | --- |
|  | 77 K | 300 K |
| $Bi_2Te_3$ | $-2.14(\pm 0.12) \times 10^{20}$ | $-3.66(\pm 0.07) \times 10^{20}$ |
| $Sb_2Te_3$ | $2.07(\pm 0.32) \times 10^{19}$ | $3.52(\pm 0.29) \times 10^{19}$ |

Table I gives the majority carrier concentrations of $Bi_2Te_3$ thin films and $Sb_2Te_3$ thin films at liquid nitrogen temperature and at room temperature (the negative values mean that the majority carrier is electron). It can be known that, when the temperature increases from 77 K to 300 K, the majority carrier (electron) concentration in $Bi_2Te_3$ thin films increases from $2.14 \times 10^{20}$ $cm^{-3}$ to $3.66 \times 10^{20}$ $cm^{-3}$, and the majority carrier (hole) concentration in $Sb_2Te_3$ thin films increases from $2.07 \times 10^{19}$ $cm^{-3}$ to $3.52 \times 10^{19}$ $cm^{-3}$.

Because the excited electrons and holes are generated in pairs by the intrinsic excitation, the majority carrier concentrations at room temperature can be considered approximately as the sum of the majority carrier concentrations at liquid nitrogen temperature and the intrinsic carrier concentrations at room temperature. The minority carrier concentrations at liquid nitrogen temperature is negligible, so it is reasonable to think that, at room temperature, the minority carrier concentrations are equal to the intrinsic carrier concentrations which are the differences between the majority carrier concentrations at liquid nitrogen temperature and that at room temperature. Therefore, from the data in Table I, it can be calculated that, at room temperature, the minority carrier concentration in $Bi_2Te_3$ is about $1.52 \times 10^{20}$ $cm^{-3}$ and that in $Sb_2Te_3$ is about $1.45 \times 10^{19}$ $cm^{-3}$. Clearly, they are comparable with the corresponding majority carrier concentrations.

It can also be known from Table I that the minority carrier (hole) concentration in n-type $Bi_2Te_3$ thin films are much higher than the majority carrier (hole) concentration in p-type $Sb_2Te_3$ thin films at room temperature. This leads to a case that both the majority carriers (electrons) and the minority carriers (holes) in $Bi_2Te_3$ thin films diffuse in the same direction from the n-type $Bi_2Te_3$ to the p-type $Sb_2Te_3$, as shown in Fig. 4. This is different from the bi-directional diffusion in a typical p-n junction.

Because, in n-type $Bi_2Te_3$ thin films, the electron's concentration is higher than the hole's concentration, the diffusion results in a net accumulation of the electrons in the p-type region. This results in a built-in electric field near the interface, which prevents the diffusion of the electrons and promote the diffusion of the holes. With the more electrons entering into the p-type region, the built-in electric field becomes stronger, so the diffusion of the electrons is gradually weakened until the diffusion of the electrons is completely balanced by the diffusion of the holes.

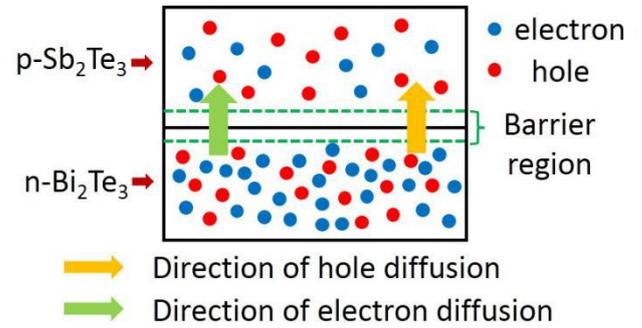

FIG. 4. Diffusion of the carriers in $Bi_2Te_3/Sb_2Te_3$ heterojunctions at room temperature (300 K).

Different from the typical p-n junction in which the intrinsic excitation is ignorable, the intrinsic excitation play an important role during the formation of $Bi_2Te_3/Sb_2Te_3$ heterojunction. Driven by the built-in electric field, the intrinsically excited electrons and holes near the interface drift to the n-type region and the p-type region, respectively. These drifting carriers neutralize the accumulated charges which come from the carrier's diffusion and weaken the built-in electric field. So, when a voltage is applied on the heterojunctions, they behave like a resistor rather than a p-n junction. This is the reason why the I-V curve in Fig. 3(b) is similar to a straight line instead of a typical I-V curve of a p-n junction.

The thermophotovoltaic cells including $Bi_2Te_3/Sb_2Te_3$ heterojunctions were tested at room temperature. Figure 5 shows the short-circuit current and the open-circuit voltage under the radiation from the heat source with different temperature. It can be seen that, when the temperature of the heat source rises from 300 K to 470 K, the short-circuit current increases from 0.055 μA to 1.2 μA and the open-circuit voltage increases from 2.6 μV to 57.25 μV. The rise in the short-circuit currents results from the increasing number of the emitted infrared photons when the temperature of the heat source increases. The rise in the open-circuit voltages is because the average energy of the emitted infrared photons increases with the temperature of the heat source.

It can be known from Fig. 5 that $Bi_2Te_3/Sb_2Te_3$ thermophotovoltaic cells have small output electrical power. The three possible reasons are responsible for this: (1) the weak radiation from the heat source; (2) the heterojunction's energy structure which results in the low open-circuit voltages; (3) the carrier recombination which lowers the short-circuit currents and the open-circuit voltages.

The radiation from the heat source can be approximately seen as blackbody radiation whose energy is weak and depends on the temperature of the heat source. The small number of the emitted infrared photons makes the short-circuit current small and the low average energy of the photons partially explain the low open-circuit voltage.

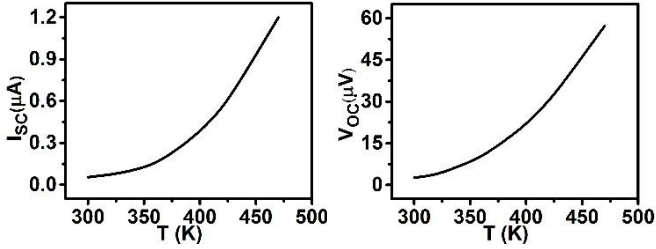

FIG. 5. Variation of short-circuit current (left) and open-circuit voltage (right) of the $Bi_2Te_3/Sb_2Te_3$ thermophotovoltaic cells with the temperature of the heat source.

The main reason for the low open-circuit voltage of the cells is the energy structures of $Bi_2Te_3$ and $Sb_2Te_3$. According to semiconductor theory [18], the Fermi level $E_{F1}$ of n-type $Bi_2Te_3$ films should satisfies

$$n_0 = n_{i1}\exp\left[(E_{F1} - E_{i1})/(K_B T)\right] \quad (1)$$

where $n_0$ is the electron concentration in the conduction band, $n_{i1}$ is the intrinsic carrier concentration, $E_{i1}$ is the intrinsic energy level, $K_B$ is Boltzmann constant, $T$ is absolute temperature. Here, as mentioned above, the difference between the electron concentrations at liquid nitrogen temperature and at room temperature (shown in Table I) may roughly be considered as the intrinsic carrier concentration, and the majority carrier concentration can be roughly considered as the electron concentration in the conduction band. By introducing these values into Eq (1), the Fermi level $E_{F1}$ can be written as

$$E_{F1} = E_{i1} + \alpha, \quad (2)$$

where the value of $\alpha$ is in the range from 0.02028 to 0.02582 eV.

Similarly, for p-type $Sb_2Te_3$ films, the Fermi level $E_{F2}$ is

$$E_{F2} = E_{i2} - \beta, \quad (3)$$

where $E_{i2}$ is the intrinsic energy level of $Sb_2Te_3$ films, and the value of $\beta$ is in the range from 0.01593 to 0.03502 eV.

Therefore, at room temperature, for $Bi_2Te_3/Sb_2Te_3$ heterojunction, the difference between two Fermi Levels can be written as:

$$E_{F1} - E_{F2} = E_{i1} - E_{i2} + \alpha + \beta, \quad (4)$$

Because the sum of $\alpha$ and $\beta$ is in the range of 36.21-60.84 meV, the value of $E_{F1} - E_{F2}$ depends mainly on the difference between two intrinsic energy levels $E_{i1} - E_{i2}$.

From previous references [19-26], it can be known that the electron affinity of $Bi_2Te_3$ is in the range of 4.125-4.525 eV [19], the electron affinity of $Sb_2Te_3$ is 4.15 eV [20], the bandgap of $Bi_2Te_3$ is 0.15-0.17 eV [21-23] and the bandgap of $Sb_2Te_3$ is 0.20-0.22 eV [24-26]. According to Eq. (4), there is a possibility that the intrinsic energy level of $Bi_2Te_3$ is lower than that of $Sb_2Te_3$. This might leads to the case that the Fermi level $E_{F1}$ of $Bi_2Te_3$ is slightly higher than $E_{F2}$ of $Sb_2Te_3$, as shown in Fig. 6. In principle, the value of the open-circuit voltage (unit is V) of the cells should not be higher than that of the difference between two Fermi levels (unit is eV) $E_{F1} - E_{F2}$ [27]. Therefore, the small value of $E_{F1} - E_{F2}$ determines that the open-circuit voltage of the $Bi_2Te_3/Sb_2Te_3$ thermophotovoltaic cells is low.

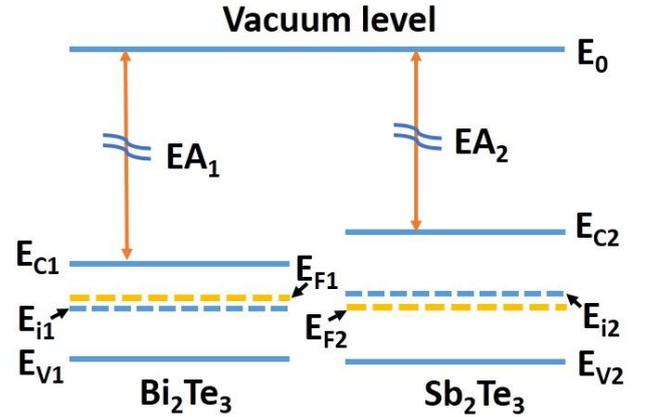

FIG. 6. Band diagrams of $Bi_2Te_3$ and $Sb_2Te_3$ at 300 K. $E_0$ is vacuum energy level, $EA_1$, $E_{C1}$, $E_{F1}$, $E_{i1}$ and $E_{V1}$ are the electron affinity, the bottom of the conduction band, the Fermi level, the intrinsic energy level and the top of the valence band of $Bi_2Te_3$, respectively. $EA_2$, $E_{C2}$, $E_{i2}$, $E_{F2}$ and $E_{V2}$ are the electron affinity, the bottom of the conduction band, the intrinsic energy level, the Fermi level, and the top of the valence band of $Sb_2Te_3$, respectively.

Another possible reason for the small output electrical power is the carrier recombination. Because both $Bi_2Te_3$ and $Sb_2Te_3$ thin films are polycrystalline, in which there exist a lot of defects such as grain boundaries and vacancies, the carriers can easily be recombined. This lowers the short-circuit currents and the open-circuit voltages [28, 29].

Although the output electrical power of $Bi_2Te_3/Sb_2Te_3$ thermophotovoltaic cells is small, the obtained results are still of significance because it has been demonstrated that the electricity can be generated by the narrow bandgap semiconductor thermophotovoltaic cells through absorbing the infrared photons emitted from the room-temperature surroundings. Compared with photovoltaic cells and high-temperature thermophotovoltaic cells, thermophotovoltaic cells in this work have more extensive applications because of their low requirement for the radiation sources.

To improve the performance of the cells, one of possible solutions is to make the p-n junctions by two different semiconductors (at least one has a narrow bandgap) whose Fermi levels have a bigger difference so that the open-circuit voltage could be enhanced.

## IV. CONCLUSIONS

In $Bi_2Te_3$/$Sb_2Te_3$ heterojunction，the intrinsic excitation makes the electrons and holes diffuse in the same direction from n-type $Bi_2Te_3$ to p-type $Sb_2Te_3$ instead of conventional bi-directional diffusion. The narrow bandgap $Bi_2Te_3$/$Sb_2Te_3$ thermophotovoltaic cells can output electrical power by absorbing the radiation from the heat source with the temperature from 300 K to 470 K. The weak radiation from the heat source, the energy structures of $Bi_2Te_3$/$Sb_2Te_3$ thermophotovoltaic cells and the carriers' recombination are responsible for the low output electrical power of the cells. The significance of this work is that it verifies that the electricity can be generated by the narrow bandgap semiconductor thermophotovoltaic cells through absorbing the radiation from dark and room-temperature surroundings.

## ACKNOWLEDGMENTS

The authors thank the supports from the Natural Science Foundation of Jiangsu (Grant No. BK2011033) and the Primary Research & Development Plan of Jiangsu Province (Grant No. BE2016175).

## REFERENCES

[1] H. Daneshvar, R. Prinja, and N. P. Kherani, Thermophotovoltaics: fundamentals, challenges and prospects, Appl. Energ. **159**, 560 (2015).

[2] B. C. Juang, R. B. Laghumavarapu, B. J. Foggo, P. J. Simmonds, A. Lin, B. L. Liang, and D. L. Huffaker, GaSb thermophotovoltaic cells grown on GaAs by molecular beam epitaxy using interfacial misfit arrays, Appl. Phys. Lett. **106**, 111101 (2015).

[3] Q. Lu, X. Zhou, A. Krysa, A. Marshall, P. Carrington, C. H. Tan, and A. Krier, InAs thermophotovoltaic cells with high quantum efficiency for waste heat recovery applications below 1000℃, Sol. Energ. Mat. Sol. C. **179**, 334 (2017).

[4] K. Qiu, and A. C. S. Hayden, Direct thermal to electrical energy conversion using very low bandgap TPV cells in a gas-fired furnace system, Energ. Convers. Manage. **79**, 54 (2014).

[5] P. Jurczak, A. Onno, K. Sablon, and H. Liu, Efficiency of GaInAs thermophotovoltaic cells: the effects of incident radiation, light trapping and recombinations, Opt. Express **23**, A1208 (2015).

[6] R. S. Tuley, J. M. S. Orr, R. J. Nicholas, D. C. Rogers, P. J. Cannard, and S. Dosanjh, Lattice-matched InGaAs on InP thermophovoltaic cells, Semicond. Sci. Tech. **28**, 015013 (2013).

[7] L. Tang, C. Xu, Z. Liu, Q. Lu, A. Marshall, and A. Krier, Suppression of the surface "dead region" for fabrication of GaInAsSb thermophotovoltaic cells, Sol. Energ. Mat. Sol. C. **163**, 263 (2017).

[8] Y. Y. Lou, X. L. Zhang, A. B. Huang, and Y. Wang, Enhanced thermal radiation conversion in a GaSb/GaInAsSb tandem thermophotovoltaic cell, Sol. Energ. Mat. Sol. C. **172**, 124 (2017).

[9] H. Lotfi, L. Li, L. Lei, R. Q. Yang, J. F. Klem, and M. B. Johnson, Narrow-bandgap interband cascade thermophotovoltaic cells, IEEE J. Photovolt. **7**, 1462 (2017).

[10] J. Yin, and R. Paiella, Limiting performance analysis of cascaded interband/intersubband thermophotovoltaic devices, Appl. Phys. Lett. **98**, 041103 (2011).

[11] T. J. Bright, L. P. Wang, and Z. M. Zhang, Performance of near-field thermophotovoltaic cells enhanced with a backside reflector, J. Heat Trans. **136**, 062701 (2014).

[12] A. Krier, M. Yin, A. R. J. Marshall, and S. E. Krier, Low bandgap InAs-based thermophotovoltaic cells for heat-electricity conversion, J. Electron. Mater. **45**, 2826 (2016).

[13] M. Tan, L. Ji, Y. Wu, P. Dai, Q. Wang, K. Li, T. Yu, Y. Yu, S. Lu, and H. Yang, Investigation of InGaAs thermophotovoltaic cells under blackbody radiation, Appl. Phys. Express **7**, 096601 (2014).

[14] W. R. Chan, P. Bermel, R. C. N. Pilawa-Podgurski, C. H. Marton, K. F. Jensen, J. J. Senkevich, J. D. Joannopoulos, M. Soljacic and I. Celanovic, Toward high-energy-density, high-efficiency, and moderate-temperature chip-scale thermophotovoltaics, P. Natl. Acad. Sci. USA. **110**, 5309 (2013).

[15] L. Tang, L. M. Fraas, Z. Liu, C. Xu, and X. Chen, Performance improvement of the GaSb thermophotovoltaic cells with n-type emitters, IEEE T. Electron Dev. **62**, 2809 (2015).

[16] K. Seeger, *Semiconductor physics* (Springer, New York, 2004) p. 43.

[17] A. Rogalski, Infrared detectors: status and trends, Prog. Quant. Electron. **27**, 59 (2003).

[18] R. H. Kingston, and S. F. Neustadter, Calculation of the space charge, electric field, and free carrier concentration at the surface of a semiconductor, J. Appl. Phys. **26**, 718 (1955).

[19] J. Nagao, E. Hatta, and K. Mukasa, in *15th International Conference on Thermoelectrics, Pasadena*, *1996*, p. 404.

[20] M. A. Islam, Y. Sulaiman, and N. Amin, A comparative study of BSF layers for ultra-thin CdS: O/CdTe solar cells, Chalcogenide Lett. **8**, 65 (2011).

[21] L. Yang, Z. G. Chen, M. Hong, G. Han, and J. Zou, Enhanced thermoelectric performance of nanostructured $Bi_2Te_3$ through significant phonon scattering, ACS Appl. Mater. Inter. **7**, 23694 (2015).

[22] T. C. Harman, B. Paris, S. E. Miller, and H. L. Goering, Preparation and some physical properties of $Bi_2Te_3$, $Sb_2Te_3$,

and As$_2$Te$_3$, J. Phys. Chem. Solids **2**, 181 (1957).

[23] D. O. Scanlon, P. D. C. King, R. P. Singh, A. De La Torre, S. M. Walker, G. Balakrishnan, F. Baumberger, and C. R. A. Catlow, Controlling bulk conductivity in topological insulators: key role of anti−site defects, Adv. Mater. **24**, 2154 (2012).

[24] T. A. Nguyen, D. Backes, A. Singh, R. Mansell, C. Barnes, D. A. Ritchie, G. Mussler, M. Lanius, D. Grutzmacher, and V. Narayan, Topological states and phase transitions in Sb$_2$Te$_3$-GeTe multilayers, Sci. Rep. **6**, 27716 (2016).

[25] I. Lefebvre, M. Lannoo, G. Allan, A. Ibanez, J. Fourcade, J. C. Jumas, and E. Beaurepaire, Electronic properties of antimony chalcogenides, Phys. Rev. Lett. **59**, 2471 (1987).

[26] A. Lawal, A. Shaari, R. Ahmed, and N. Jarkoni, Sb$_2$Te$_3$ crystal a potential absorber material for broadband photodetector: A first-principles study, Results Phys. **7**, 2302 (2017).

[27] J. P. Ponpon, and P. Siffert, Open-circuit voltage of MIS silicon solar cells, J. Appl. Phys. **47**, 3248 (1976).

[28] J. Chantana, T. Kato, H. Sugimoto, and T. Minemoto, Investigation of correlation between open-circuit voltage deficit and carrier recombination rates in Cu(In,Ga)(S,Se)$_2$-based thin-film solar cells, Appl. Phys. Lett. **112**, 151601 (2018).

[29] N. D. Marco, H. Zhou, Q. Chen, P. Sun, Z. Liu, L. Meng, E. Yao, Y. Liu, A. Schiffer, and Y. Yang, Guanidinium: A route to enhanced carrier lifetime and open-circuit voltage in hybrid perovskite solar cells, Nano Lett. **16**, 1009 (2016).